\documentclass[showpacs,preprintnumbers,showkeys,superscriptaddress,twocolumn]{revtex4}  
\usepackage{amssymb}
\usepackage{graphicx}
\usepackage{dcolumn}
\usepackage{bm}
\usepackage{appendix}
\usepackage{epstopdf}

\def\eqref#1{Eq.~(\ref{eq:#1})}

\begin{document}

\title{Application of the variational principle to a coherent-pair condensate: The HFB case}
\author{L. Y. Jia}  \email{liyuan.jia@usst.edu.cn}
\affiliation{Department of Physics, University of Shanghai for
Science and Technology, Shanghai 200093, P. R. China}

\date{\today}


\begin{abstract}

Recently we proposed a scheme that applies the variational principle to a coherent-pair condensate in the BCS case \cite{Jia_2019}. This work extends the scheme to the HFB case by allowing variation of the canonical single-particle basis. The result is equivalent to that of the so-called variation after particle-number projection in the HFB case, but now the particle number is always conserved and the time-consuming projection is avoided. Specifically, we derive the analytical expression for the gradient of the average energy with respect to changes of the canonical basis, which is then used in the family of gradient minimizers to iteratively minimize energy. In practice, we find the so-called ADAM minimizer (adaptive moments of gradient), borrowed from the machine-learning field, is very effective. The new algorithm is demonstrated in a semi-realistic example using the realistic $V_{{\rm{low}}{\textrm{-}}k}$ interaction and large model spaces (up to $15$ harmonic-oscillator major shells). It easily runs on a laptop, and practically the computer time cost (in the HFB case) is several times that of solving HF by the gradient minimizers. We hope the new algorithm could become a common practice because of its simplicity.

\end{abstract}


\vspace{0.4in}

\maketitle


\section{Introduction}

The self-consistent mean-field theory is widely used in nuclear structure \cite{Bender_2003}. Its simplicity allows application to the whole nuclear chart, and the self-consistent single-particle levels are the starting point for nuclear shell model (configuration interaction), especially for nuclei far from stability where the evolution of single-particle levels is unknown a priori. The Hartree-Fock (HF) method is the microscopic mean-field theory.

The collective (coherent) pairing effect \cite{Bohr_1958} is important across the nuclear chart \cite{Bohr_book, Ring_book}. To incorporate pairing into the mean field, one introduces quasiparticles and the microscopic theory is the Hartree-Fock-Bogoliubov (HFB) method \cite{Belyaev_1959}. HFB is simply the variational principle using the quasiparticle vacuum as the trial wavefunction \cite{Ring_book}. Quasiparticles are highly successful and wildly used. As fermions, they have clear physical picture and allow easy computation. However, quasiparticles break the exact particle number \cite{Ring_book}. Only the average particle number is guaranteed by the chemical potential; effectively, one replaces the target nucleus by an average of neighbouring nuclei. This average may cause errors, especially in phase transition regions of sharp property changes.

To cue the problems, one projects the HFB quasiparticle vacuum onto good particle number; the standard way to project is by numerical integration over the gauge angle \cite{Dietrich_1964, Ring_book}. The projection can be done after or before the variation \cite{Ring_book, Bender_2003}. For examples of projection after variation (PAV), see Refs. \cite{Ring_book, Bender_2003} and references therein. For examples of variation after projection (VAP), see Refs. \cite{Dietrich_1964, Dussel_2007, Dukelsky_2000, Sandulescu_2008} in the BCS case and Refs. \cite{Sheikh_2000, Anguiano_2001, Anguiano_2002, Stoitsov_2007, Hupin_2012} in the HFB case. When feasible, VAP is preferred \cite{Ring_book, Jia_2019, Bender_2003} over PAV. The practical difficulty of VAP is that numerical projection by integration is time-consuming \cite{Wang_2014} and needed many times when performing VAP. In the literature there are far fewer realistic applications of VAP+HFB than those of the HFB theory without projection.


It is easy to analytically project \cite{Ring_book} the HFB quasiparticle vacuum onto good particle number --- the result is a coherent-pair condensate [see Eq. (\ref{gs})]. This work applies the variational principle directly to this coherent-pair condensate (VDPC). The particle number is always conserved and the time-consuming projection is avoided. (This feature is emphasized by the word ``directly'' in the name ``VDPC''. We name the new method VDPC because VAP may be misleading: there is no projection at all.) The variational parameters are the canonical single-particle basis and the coherent-pair structure $v_\alpha$ on this basis. Ref. \cite{Jia_2019} has proposed a fast algorithm for varying $v_\alpha$ at fixed canonical basis, which is VDPC+BCS and the result is equivalent to that of VAP+BCS. This work considers VDPC+HFB that varies $v_\alpha$ and the canonical basis together, and the result is equivalent to that of VAP+HFB.


Specifically, we first consider the subproblem of varying the canonical basis at fixed $v_\alpha$, which we call VDPC+CB (plus canonical basis). This generalizes the HF theory, which uses special values of $v_\alpha$: $v_\alpha = 1$ for occupied levels in the HF Slater determinant and $v_\alpha = 0$ for empty levels. To solve VDPC+CB, we derive the analytical expression for the gradient of the average energy with respect to changes of the canonical basis, which is then used in the family of gradient minimizers \cite{Reinhard_1982, Robledo_2011, Ring_book} to iteratively minimize energy. In practice, we find the so-called ADAM minimizer (adaptive moments of gradient) \cite{Adam_2014}, borrowed from the machine-learning field \cite{Goodfellow_book}, is very effective (energy converges quicker than other minimizers).

The new VDPC+HFB algorithm combines VDPC+BCS and VDPC+CB. Specifically, we insert varying $v_\alpha$ (VDPC+BCS) into several places in the process of varying the canonical basis (VDPC+CB). (It is unnecessary to vary $v_\alpha$ after every iteration of VDPC+CB; after every $20$ iterations, for example, is enough.)

We demonstrate the new VDPC+CB and VDPC+HFB algorithms in a semi-realistic example using the realistic $V_{{\rm{low}}{\textrm{-}}k}$ interaction \cite{Bogner_2003} and large model spaces (up to $15$ harmonic-oscillator major shells). (The same example/Hamiltonian has been used in Ref. \cite{Jia_2019} for VDPC+BCS.) The energy-convergence pattern and actual computer time cost are given in detail. The new algorithms easily run on a laptop, and practically the computer time cost to solve VDPC+HFB (time to solve VDPC+CB) is about $4 \sim 6$ times ($2 \sim 3$ times) that to solve HF by the gradient minimizers. We hope the new VDPC+HFB algorithm could become a common practice because of its simplicity.


This work relates to Refs. \cite{Dukelsky_2000, Hupin_2011, Sandulescu_2008, Hupin_2012, Hupin_2011_2}. The average energy of the coherent-pair condensate (\ref{gs}) has been derived for the pairing Hamiltonian \cite{Dukelsky_2000} and a general Hamiltonian \cite{Hupin_2011}. However, the gradient of energy with respect to $v_\alpha$ and to changes of the canonical basis have not been derived. (The gradient with respect to $v_\alpha$ was recently derived in Ref. \cite{Jia_2019}.) For successes and limitations of these works \cite{Dukelsky_2000, Hupin_2011, Sandulescu_2008, Hupin_2012, Hupin_2011_2}, please see the introduction part of Ref. \cite{Jia_2019}. For modern mean-field theories using large model spaces in the pairing channel, VAP has been done only by the numerical gauge-angle integration \cite{Anguiano_2001, Anguiano_2002, Stoitsov_2007}. This work aims to propose the new VDPC algorithm.

We should also mention the Lipkin-Nogami prescription to restore the particle number approximately \cite{Lipkin_1960, Nogami_1964, Pradhan_1973}. It is widely used because the exact VAP (by the numerical gauge-angle integration) is computationally expensive \cite{Bender_2003, Wang_2014}. Ongoing efforts exist to improve the Lipkin method \cite{Wang_2014}.

This work is organized as follows. Section \ref{Sec_PC} briefly reviews
the formalism for the coherent-pair condensate --- the trial wavefunction. Section \ref{Sec_gradient} derives analytically the gradient of the average energy with respect to changes of the canonical single-particle basis. What this gradient expression becomes when reducing the valence space in the pairing channel (make it smaller than the full space) is discussed in Sec. \ref{Sec_limit}. We explain how to solve VDPC+CB by the family of iterative gradient minimizers in Sec. \ref{Sec_VDPC_CB}, and describe the VDPC+HFB algorithm in Sec. \ref{Sec_VDPC_HFB}. Section \ref{Sec_example} applies the new VDPC algorithms to a semi-realistic example. Section \ref{Sec_summary} summarizes the work.


\section{Coherent-Pair Condensate as Trial Wavefunction  \label{Sec_PC}}

This work uses the coherent-pair condensate as the trial wavefunction in the variational principle. This section briefly reviews the relevant definitions and formulas. For clarity we consider one kind of nucleon, the extension to active protons and neutrons is simple: the existence of protons simply provides a correction to the neutron single-particle energy, through the two-body proton-neutron interaction. We assume time-reversal self-consistent symmetry \cite{Ring_book, Goodman_1979}, and the single-particle basis state $|\alpha\rangle$ is Kramers degenerate with its time-reversed partner $|\tilde{\alpha}\rangle$ ($|\tilde{\tilde{\alpha}}\rangle = - |\alpha\rangle$). No other symmetry is assumed.

For the ground state of the $2N$-particle system, the trial wavefunction is an $N$-pair condensate,
\begin{eqnarray}
|\phi_N\rangle = \frac{1}{\sqrt{\chi_{N}}} (P^\dagger)^{N} |0\rangle
, \label{gs}
\end{eqnarray}
where
\begin{eqnarray}
\chi_{N} = \langle0| P^N (P^\dagger)^{N} |0\rangle  \label{chi_N}
\end{eqnarray}
is the normalization factor, and $P^\dagger$ creates a coherent pair
\begin{eqnarray}
P^\dagger = \frac{1}{2} \sum_{\alpha \beta} v_{\alpha\beta}
a_\alpha^\dagger a_{\tilde{\beta}}^\dagger . \label{P_dag_alpha_beta}
\end{eqnarray}
Requiring $|\phi_N\rangle$ to be time-even implies that the pair
structure matrix $v_{\alpha\beta} = v_{\beta\alpha}^*$ is Hermitian (because $\hat{T}
v_{\alpha\beta} a_\alpha^\dagger a_{\tilde{\beta}}^\dagger |0\rangle
= v_{\alpha\beta}^* a_{\tilde{\alpha}}^\dagger a_{\tilde{\tilde{\beta}}}^\dagger
|0\rangle = v_{\alpha\beta}^* a_{\beta}^\dagger a_{\tilde{\alpha}}^\dagger
|0\rangle$). The unitary transformation to the canonical single-particle basis diagonalizes $v_{\alpha\beta}$. Below we always use the canonical basis, thus $v_{\alpha\beta} = \delta_{\alpha\beta} v_\alpha$, $v_\alpha$ is real, and the coherent pair (\ref{P_dag_alpha_beta}) becomes
\begin{eqnarray}
P^\dagger = \frac{1}{2} \sum_{\alpha} v_{\alpha}
a_\alpha^\dagger a_{\tilde{\alpha}}^\dagger = \sum_{\alpha \in \Theta} v_{\alpha}
P^\dagger_\alpha ,  \label{P_dag}
\end{eqnarray}
where
\begin{eqnarray}
P^\dagger_\alpha = a_\alpha^\dagger a_{\tilde{\alpha}}^\dagger = P^\dagger_{\tilde{\alpha}}   \label{P_dag_alpha}
\end{eqnarray}
creates a pair on $|\alpha\rangle$ and $|\tilde{\alpha}\rangle$.
In Eq. (\ref{P_dag}), $\Theta$ is the set of pair-indices that picks only one from each degenerate pair $|\alpha\rangle$ and $|\tilde{\alpha}\rangle$. For example, with axial symmetry we can choose $\Theta$ to be orbits of a positive magnetic quantum number. $\sum_{\alpha}$ and $\sum_{\alpha \in \Theta}$ means summing over single-particle indices and pair indices.


The trial wavefunction (\ref{gs}) is specified by two sets of variational parameters: the canonical single-particle basis and the pair structure $v_\alpha$ (\ref{P_dag}) on this basis. Ref. \cite{Jia_2019} has proposed a fast algorithm for varying $v_\alpha$ at fixed canonical basis that is VDPC+BCS. This work first considers varying the canonical basis at fixed $v_\alpha$ in Sec. \ref{Sec_VDPC_CB}, which is VDPC+CB. Then Sec. \ref{Sec_VDPC_HFB} considers varying the canonical basis and $v_\alpha$ together that is VDPC+HFB.

If one insists time-reversal self-consistent symmetry \cite{Ring_book, Goodman_1979}, and analytically projects the time-even HFB quasiparticle vacuum onto good particle number, one gets the same trial wavefunction (\ref{gs}) (see Eq. (7.18) of Ref. \cite{Ring_book}). Therefore, the result of VDPC+HFB is equivalent to that of VAP+HFB, and the result of VDPC+BCS \cite{Jia_2019} is equivalent to that of VAP+BCS. The difference is that VDPC always conserves the particle number and avoids the time-consuming numerical projection (gauge-angle integration) of VAP.

The coherent-pair condensate (\ref{gs}) includes the HF Slater determinant as a special case, when $v_\alpha$ is fixed to $1$ for the $2N$ occupied HF orbits and to $0$ for other empty HF orbits. From this perspective, VDPC extends the HF theory to incorporate pairing correlations, but not by introducing quasi-particles that break particle number (what HFB does); instead, VDPC always converses the particle number.


For convenience, we introduce $\{[\gamma_1 \gamma_2 \ldots \gamma_r]\}$ to represent a subspace of the original single-particle space, by removing Kramers pairs of single-particle levels $\gamma_1, \tilde{\gamma}_1, \gamma_2, \tilde{\gamma}_2, \ldots \gamma_r, \tilde{\gamma}_r$ from the latter. Later we will express the gradient of energy in terms of Pauli-blocked normalizations \cite{Jia_2017},
\begin{eqnarray}
\chi_{N}^{[\gamma_1 \gamma_2 \ldots \gamma_r]} \equiv \langle0| P^N P_{\gamma_1} P_{\gamma_2} \ldots P_{\gamma_r} P_{\gamma_1}^\dagger P_{\gamma_2}^\dagger \ldots P_{\gamma_r}^\dagger (P^\dagger)^{N} |0\rangle ,  \label{chi_N_blocked}
\end{eqnarray}
which is the normalization in the blocked subspace $\{[\gamma_1 \gamma_2 \ldots \gamma_r]\}$. Given $v_\alpha$, how to compute $\chi_{N}^{[\gamma_1 \gamma_2 \ldots \gamma_r]}$? It is explained in detail in Ref. \cite{Jia_2019}. For example, $\chi_N$ and $\chi_{N}^{[\alpha]}$ are computed by recursive relations,
\begin{eqnarray}
\chi_N = N \sum_{\alpha \in \Theta} (v_\alpha)^2 \chi_{N-1}^{[\alpha]} ,  \label{chi_rec1} \\
\chi_{N} - \chi_{N}^{[\alpha]} = (N v_\alpha)^2 \chi_{N-1}^{[\alpha]} = \chi_N \langle \phi_N | \hat{n}_\alpha | \phi_N \rangle ,  \label{n_ave}
\end{eqnarray}
with initial value $\chi_{N=0}^{[\alpha]} = 1$. Knowing $\chi_{N-1}^{[\alpha]}$'s, we compute $\chi_N$ by Eq. (\ref{chi_rec1}), and then $\chi_{N}^{[\alpha]}$'s by Eq. (\ref{n_ave}). $\langle \phi_N | \hat{n}_\alpha | \phi_N \rangle = \langle 0 | P^{N} a_\alpha^\dagger a_\alpha (P^\dagger)^{N} | 0 \rangle / \chi_N$ is the average occupation number. Equations (\ref{chi_rec1}) and (\ref{n_ave}) are also valid in the blocked subspaces $\{[\gamma_1 \gamma_2 \ldots \gamma_r]\}$, which could be used to compute $\chi_{N}^{[\alpha\beta]}$ and $\chi_{N}^{[\alpha\beta\gamma]}$; but another method is better (by Eqs. (15) and (16) of Ref. \cite{Jia_2019}). Later we will also need
\begin{eqnarray}
| \phi_N^{[\beta]} \rangle \equiv \frac{1}{\sqrt{\chi_{N}^{[\beta]}}} (P^\dagger - v_\beta P_\beta^\dagger)^{N} |0\rangle
\end{eqnarray}
that is the pair condensate with $\beta$ and $\tilde{\beta}$ blocked.

This section discusses the ``kinematics'' of the VDPC formalism, next we discuss the ``dynamics''.


\section{Gradient of Energy  \label{Sec_gradient}}

In this section we derive the gradient of the average energy with respect to changes of the canonical single-particle basis. The anti-symmetrized two-body Hamiltonian is
\begin{eqnarray}
H = \sum_{\alpha\beta} \epsilon_{\alpha\beta} a_\alpha^\dagger a_\beta + \frac{1}{4} \sum_{\alpha \beta \gamma \mu} V_{\alpha \beta \gamma \mu} a_\alpha^\dagger a_\beta^\dagger a_\gamma a_\mu . \label{H_def}
\end{eqnarray}
Note the ordering of $\alpha \beta \gamma \mu$, thus $V_{\alpha \beta \gamma \mu} = - \langle \alpha \beta| V | \gamma \mu\rangle$.
I assume time-even $H$ ($\epsilon_{\alpha\beta} = \epsilon_{\tilde{\beta}\tilde{\alpha}}$, $V_{\alpha \beta \gamma \mu} = V_{\tilde{\mu} \tilde{\gamma} \tilde{\beta} \tilde{\alpha} }$) and real $\epsilon_{\alpha\beta}$ and $V_{\alpha \beta \gamma \mu}$. No other symmetry of $H$ is assumed. The Hamiltonian matrix elements $\epsilon_{\alpha\beta}$ and $V_{\alpha \beta \gamma \mu}$ have been transformed to the canonical basis ($\alpha, \beta, \gamma, \mu$ run over the canonical-basis states).

The average energy of the coherent-pair condensate $\bar{E} = \langle \phi_N | H | \phi_N \rangle$ has been derived in Eq. (25) of Ref. \cite{Jia_2019},
\begin{eqnarray}
\langle \phi_N | H | \phi_N \rangle = \frac{N^2}{\chi_N} \Bigg( \sum_{\alpha \in \Theta} (2\epsilon_{\alpha\alpha} + G_{\alpha\alpha}) (v_\alpha)^2 \chi_{N-1}^{[\alpha]}  \nonumber \\
+ \sum^{\alpha \ne \beta}_{\alpha,\beta \in \Theta} G_{\alpha\beta} v_\alpha v_\beta \chi_{N-1}^{[\alpha\beta]}  \nonumber \\
+ (N-1)^2 \sum^{\alpha\ne\beta}_{\alpha,\beta \in \Theta} \Lambda_{\alpha\beta} (v_\alpha v_\beta)^2 \chi_{N-2}^{[\alpha\beta]} \Bigg) ,  \label{H_ave_1}
\end{eqnarray}
where we introduce the paring matrix elements $G_{\alpha\beta}$ and the ``monopole'' matrix elements $\Lambda_{\alpha\beta}$ as
\begin{eqnarray}
G_{\alpha\beta} \equiv V_{\alpha\tilde{\alpha}\tilde{\beta}\beta} ,  \label{G_12} \\
\Lambda_{\alpha\beta} \equiv V_{\alpha\beta\beta\alpha} + V_{\alpha\tilde{\beta}\tilde{\beta}\alpha} . \label{L_12}
\end{eqnarray}
Note $G_{\alpha\beta} = G_{\beta\alpha} = G_{\alpha\tilde{\beta}}$, $\Lambda_{\alpha\beta} = \Lambda_{\beta\alpha} = \Lambda_{\alpha\tilde{\beta}}$, and $G_{\alpha\alpha} = \Lambda_{\alpha\alpha}$.

The gradient of the average energy with respect to $v_\alpha$ (at fixed canonical basis), ${\partial \bar{E}}/{\partial v_{\alpha}} = {\partial[ \langle \phi_N | H | \phi_N \rangle ]}/{\partial v_{\alpha}}$, has been derived in Eqs. (31) and (32) of Ref. \cite{Jia_2019}. Based on this gradient, Ref. \cite{Jia_2019} proposed a fast algorithm for varying $v_\alpha$ at fixed canonical basis, which is VDPC+BCS.


Now we derive the gradient of the average energy with respect to changes of the canonical basis (at fixed $v_\alpha$). For this purpose we first parameterize changes of the single-particle basis. We have assumed the Hamiltonian matrix elements $\epsilon_{\alpha\beta}$ and $V_{\alpha\beta\gamma\mu}$ (\ref{H_def}) are real, so the eigen wavefunctions are also real. Therefore we restrict the trial wavefunction (\ref{gs}) (approximates the ground state) to be real, that is, $v_{\alpha\beta}$ in Eq. (\ref{P_dag_alpha_beta}) to be real. Thus the unitary transformation to the canonical basis (\ref{P_dag}) becomes an orthogonal transformation (a real unitary matrix). We parameterize the mixing of two single-particle basis states as (real mixing coefficients)
\begin{eqnarray}
|\alpha'\rangle = \cos \theta |\alpha\rangle + \sin \theta |\beta\rangle ,~ |\beta'\rangle = \cos \theta |\beta\rangle - \sin \theta |\alpha\rangle ,~  \label{tran_ori}
\end{eqnarray}
consequently for their time-reversal partners
\begin{eqnarray}
|\widetilde{\alpha'}\rangle = \cos \theta |\tilde{\alpha}\rangle + \sin \theta |\tilde{\beta}\rangle ,~ |\widetilde{\beta'}\rangle = \cos \theta |\tilde{\beta}\rangle - \sin \theta |\tilde{\alpha}\rangle .~  \label{tran_time}
\end{eqnarray}
$|\alpha\rangle$ and $|\beta\rangle$ belong to different Kramers pairs ($P_\alpha \ne P_{\beta}$). Mixing $|\alpha\rangle$ and $|\tilde{\alpha}\rangle$ has no effect on $P_\alpha^\dagger = a_\alpha^\dagger a_{\tilde{\alpha}}^\dagger$ (\ref{P_dag_alpha}), so $P^\dagger$ (\ref{P_dag}) and $|\phi_N\rangle$ (\ref{gs}) stay unchanged. If one requires additional self-consistent symmetry (for example, parity and axial symmetry), $|\alpha\rangle$ and $|\beta\rangle$ have the same values of the corresponding good quantum numbers (parity and angular momentum projection onto the symmetry axis). For infinitesimal mixing (infinitesimal $\theta \approx 0$), keeping the first-order in $\theta$, Eqs. (\ref{tran_ori}) and (\ref{tran_time}) imply variations of the basis states,
\begin{eqnarray}
\delta |\alpha\rangle = |\alpha'\rangle - |\alpha\rangle \approx \theta |\beta\rangle ,~ \delta |\beta\rangle = |\beta'\rangle - |\beta\rangle \approx - \theta |\alpha\rangle ,~  \label{d_ori} \\
\delta |\tilde{\alpha}\rangle = |\widetilde{\alpha'}\rangle - |\tilde{\alpha}\rangle \approx \theta |\tilde{\beta}\rangle ,~ \delta |\tilde{\beta}\rangle = |\widetilde{\beta'}\rangle - |\tilde{\beta}\rangle \approx - \theta |\tilde{\alpha}\rangle .~~  \label{d_time}
\end{eqnarray}


Variation of the average energy $\bar{E}$ (\ref{H_ave_1}) under an infinitesimal $\theta$ has been derived in Sec. V of the arXiv manuscript \cite{Jia_arXiv_2018} (but has not been published in a journal yet). We repeat the result here: 
\begin{eqnarray}
\delta \bar{E} = \delta (\langle \phi_N | H | \phi_N \rangle) = 4 \theta f_{\alpha\beta} ,  \label{E_var}
\end{eqnarray}
where
\begin{widetext}
\begin{eqnarray}
f_{\alpha\beta} = \frac{N^2 (v_\alpha - v_\beta)}{\chi_N } \Bigg( [ (v_\alpha + v_\beta) \epsilon_{\alpha\beta} + v_\alpha V_{\alpha\tilde{\beta}\tilde{\alpha}\alpha}
+ v_\beta V_{\alpha\tilde{\beta}\tilde{\beta}\beta} ] \chi_{N-1}^{[\alpha\beta]}  \nonumber \\
+ \sum_{\gamma \in \Theta}^{P_\gamma \ne P_\alpha, P_\beta} v_\gamma V_{\alpha\tilde{\beta}\tilde{\gamma}\gamma} [ \chi_{N-1}^{[\alpha\beta\gamma]} - (N-1)^2 v_\alpha v_\beta \chi_{N-2}^{[\alpha\beta\gamma]}]  \nonumber \\
+ (N-1)^2 (v_\alpha + v_\beta) \sum_{\gamma \in \Theta}^{P_\gamma \ne P_\alpha, P_\beta} (v_\gamma)^2 ( V_{\alpha\gamma\gamma\beta} + V_{\alpha\tilde{\gamma}\tilde{\gamma}\beta} ) \chi_{N-2}^{[\alpha\beta\gamma]} \Bigg) .  \label{f_12}
\end{eqnarray}
$f_{\alpha\beta} = - f_{\beta\alpha}$ is skew-symmetric. We pull out the factor $4$ when defining $f_{\alpha\beta}$ in Eq. (\ref{E_var}), so that $f_{\alpha\beta}$ reduces to the off-diagonal part of the HF mean field when $|\phi_N\rangle$ (\ref{gs}) reduces to a HF Slater determinant, as shown in Sec. \ref{Sec_limit}. Equation (\ref{E_var}) means that $4 f_{\alpha\beta}$ is the partial derivative at $\theta = 0$,
\begin{eqnarray}
\frac{\partial \bar{E}}{\partial \theta} |_{\theta = 0} = \frac{\partial (\langle \phi_N | H | \phi_N \rangle)}{\partial \theta} |_{\theta = 0} = 4 f_{\alpha\beta} ,  \label{partialE}
\end{eqnarray}
where $\theta$ is the angle mixes the two canonical basis states $|\alpha\rangle$ and $|\beta\rangle$ according to Eq. (\ref{tran_ori}). Using $(N-1)^2 (v_\gamma)^2 \chi_{N-2}^{[\alpha\beta\gamma]} = \chi_{N-1}^{[\alpha\beta]} \langle \phi_{N-1}^{[\alpha\beta]} | \hat{n}_\gamma | \phi_{N-1}^{[\alpha\beta]} \rangle$
[Eq. (\ref{n_ave}) with $N \rightarrow N-1$, $\alpha \rightarrow \gamma$, then blocking the $\alpha$ and $\beta$ Kramers pairs], an equivalent form of $f_{\alpha\beta}$ is
\begin{eqnarray}
f_{\alpha\beta} = \frac{N^2 (v_\alpha - v_\beta)}{\chi_N }    \Bigg( [ (v_\alpha + v_\beta) (\epsilon_{\alpha\beta} + \sum_{\gamma}^{P_\gamma \ne P_\alpha, P_\beta} V_{\alpha\gamma\gamma\beta} \langle \phi_{N-1}^{[\alpha\beta]} | \hat{n}_\gamma | \phi_{N-1}^{[\alpha\beta]} \rangle ) + v_\alpha V_{\alpha\tilde{\beta}\tilde{\alpha}\alpha}
+ v_\beta V_{\alpha\tilde{\beta}\tilde{\beta}\beta} ] \chi_{N-1}^{[\alpha\beta]}  \nonumber \\
+ \frac{1}{2} \sum_{\gamma}^{P_\gamma \ne P_\alpha, P_\beta} v_\gamma V_{\alpha\tilde{\beta}\tilde{\gamma}\gamma} [ \chi_{N-1}^{[\alpha\beta\gamma]} - (N-1)^2 v_\alpha v_\beta \chi_{N-2}^{[\alpha\beta\gamma]}]
 \Bigg) ,  \label{f_12_2}
\end{eqnarray}
\end{widetext}
where $\gamma$ sums over single-particle index. In Eq. (\ref{f_12_2}), the $\sum_{\gamma} V_{\alpha\gamma\gamma\beta} \langle \phi_{N-1}^{[\alpha\beta]} | \hat{n}_\gamma | \phi_{N-1}^{[\alpha\beta]} \rangle$ term is the monopole correction to $\epsilon_{\alpha\beta}$. 

In this section, we derive the gradient (\ref{f_12}) when assuming all the single-particle levels (canonical basis states) participate in the coherent pairing (\ref{P_dag}). Usually, only the levels near the Fermi surface are important, and they compose the active valence space in the pairing channel (smaller dimension than the full space). In the next section, we consider what the gradient (\ref{f_12}) becomes when using such a limited valence space in the pairing channel.


\section{Reducing Valence Space in Pairing Channel  \label{Sec_limit}}

In the previous section, we derive the energy gradient expression (\ref{f_12}) when assuming all the single-particle levels (canonical basis states) participate in the coherent pairing (\ref{P_dag}). In this section, we consider what this gradient expression becomes when reducing the valence space in the pairing channel.

Usually, only the single-particle levels (canonical basis states) near the Fermi energy $E_F$  actively participate in the coherent pairing (\ref{P_dag}) and are partially filled. The levels well below and well above $E_F$ are almost full and empty ($n_\alpha \approx 1$ and $0$). So it is a good approximation to keep these levels completely full and empty ($n_\alpha = 1$ and $0$), and restrict the coherent pairing (\ref{P_dag}) to those levels near the Fermi surface. The latter compose the active valence space in the pairing channel, which should be large enough (include enough levels) for a desired accuracy of the approximation.


Specifically, we divide the full single-particle space (canonical basis) into three subspaces: $L$ (low), $V$ (valence), and $H$ (high). For $\alpha \in H$, we set $v_\alpha = 0$ in Eq. (\ref{P_dag}), so $n_\alpha = 0$. For $\alpha \in L$, we set $v_\alpha = v_L$ and will take the limit $v_L \rightarrow \infty$, so $n_\alpha \rightarrow 1$. For $\alpha \in V$, $v_\alpha$ is a finite nonzero number, so $0 < n_\alpha < 1$. The criterion for division (into $L$, $V$, $H$ subspaces) is not necessarily by single-particle energy; in fact, this work will use the criterion by occupation number $n_\alpha$. We assume $L$ has dimension $2 N_L$, thus the valence particle number in $V$ is $2(N-N_L) \equiv 2 \bar{N}$. We introduce the $V$-subspace coherent pair
\begin{eqnarray}
\bar{P}^\dagger \equiv \frac{1}{2} \sum_{\alpha \in V} v_{\alpha}
P^\dagger_{\alpha} = \sum_{\alpha \in \Theta, \alpha \in V} v_{\alpha}
P^\dagger_{\alpha}  \label{P_bar_dag} ,
\end{eqnarray}
and the $V$-subspace coherent-pair condensate
\begin{eqnarray}
|\bar{\phi}_{\bar{N}}\rangle \equiv \frac{1}{\sqrt{\bar{\chi}_{\bar{N}}}} (\bar{P}^\dagger)^{\bar{N}} |0\rangle ,  \label{gs_V}
\end{eqnarray}
where
\begin{eqnarray}
\bar{\chi}_{\bar{N}} \equiv \langle0| \bar{P}^{\bar{N}} (\bar{P}^\dagger)^{\bar{N}} |0\rangle   \label{chi_bar_N}
\end{eqnarray}
is the $V$-subspace normalization. In summary, we hat the $V$-subspace symbols by a bar (not to be confused with the average energy $\bar{E} = \langle \phi_N | H | \phi_N \rangle$).

It is easy to derive the relations between the full-space quantities and the $V$-subspace ones. For the coherent pair (\ref{P_dag}),
\begin{eqnarray}
P^\dagger = \bar{P}^\dagger + v_{L} \sum_{\alpha \in \Theta, \alpha \in L}
P^\dagger_{\alpha} .  \label{P_dag_limit}
\end{eqnarray}
So for $(P^\dagger)^N = (P^\dagger)^{N_L + \bar{N}}$,
\begin{eqnarray}
(P^\dagger)^N = A_N^{N_L} (v_{L})^{N_L} (\bar{P}^\dagger)^{\bar{N}} (\prod_{\alpha \in \Theta, \alpha \in L} P_{\alpha}^\dagger)  \nonumber \\
+ O[(v_{L})^{N_L-1}] ,  \label{P_N_limit}
\end{eqnarray}
where $O[(v_{L})^{N_L-1}]$ represents terms of the power $N_L-1$ and lower. $A_N^{N_L} = N! / (N-N_L)! = N! / \bar{N}!$ is the number of permutations, for selecting the $N_L$ of $P_{\alpha \in L}^\dagger$ operators from the $N$ of multiplying $P^\dagger$. $\prod_{\alpha \in \Theta, \alpha \in L} P_{\alpha}^\dagger$ equals $\prod_{\alpha \in L} a_{\alpha}^\dagger$ within a sign that will fully occupy the $L$-subspace when acting on $|0\rangle$. Using Eq. (\ref{P_N_limit}), the normalization (\ref{chi_N}) is
\begin{eqnarray}
\chi_{N} = (A_N^{N_L})^2 (v_{L})^{2N_L} \bar{\chi}_{\bar{N}} + O[(v_{L})^{2N_L-1}] .  \label{chi_N_limit}
\end{eqnarray}
Using Eqs. (\ref{P_N_limit}) and (\ref{chi_N_limit}), the coherent-pair condensate (\ref{gs}) is
\begin{eqnarray}
|\phi_N\rangle = \frac{1}{\sqrt{\bar{\chi}_{\bar{N}}}} (\bar{P}^\dagger)^{\bar{N}} (\prod_{\alpha \in \Theta, \alpha \in L} P_{\alpha}^\dagger) |0\rangle + O[(v_{L})^{-1}]  \nonumber \\
= (\prod_{\alpha \in \Theta, \alpha \in L} P_{\alpha}^\dagger) |\bar{\phi}_{\bar{N}}\rangle + O[(v_{L})^{-1}]
. \label{gs_limit}
\end{eqnarray}
When taking the limit $v_L \rightarrow \infty$, the $O[(v_{L})^{-1}]$ terms vanish, and the trial wavefunction $|\phi_N\rangle$ becomes
\begin{eqnarray}
(\prod_{\alpha \in \Theta, \alpha \in L} P_{\alpha}^\dagger) |\bar{\phi}_{\bar{N}}\rangle , \label{gs_new}
\end{eqnarray}
which is a $V$-subspace pair condensate $|\bar{\phi}_{\bar{N}}\rangle$ (\ref{gs_V}) plus a fully occupied $L$-subspace core $\prod_{\alpha \in \Theta, \alpha \in L} P_{\alpha}^\dagger |0\rangle$ and an empty $H$-subspace. That is, we restrict (reduce) the valence space in the pairing channel to be the $V$-subspace.


Now we derive the energy gradient expression when using the trial wavefunction (\ref{gs_new}). This can be done by taking the limit $v_L \rightarrow \infty$ on Eq. (\ref{f_12}) or Eq. (\ref{f_12_2}); here we skip the five-page long derivation and only show the results in Eqs. (\ref{f_VV})-(\ref{f_VH}). For convenience, we introduce new symbols
\begin{eqnarray}
K_{\alpha\beta\gamma\mu} &\equiv& V_{\alpha\tilde{\beta}\tilde{\gamma}\mu} ,  \nonumber \\
W_{\alpha \beta \gamma \mu} &\equiv& V_{\alpha \gamma \mu \beta} + V_{\alpha \tilde{\gamma} \tilde{\mu} \beta} ,  \nonumber
\end{eqnarray}
and the HF mean field generated by the fully occupied $L$-subspace (the core)
\begin{eqnarray}
\lambda_{\alpha\beta} \equiv \epsilon_{\alpha\beta} + \sum_{\gamma \in L} V_{\alpha\gamma\gamma\beta} .  \label{HF_Lcore}
\end{eqnarray}
For mixing within the $V$-subspace,
\begin{eqnarray}
f_{\alpha \in V, \beta \in V} = \frac{\bar{N}^2 (v_\alpha - v_\beta)}{\bar{\chi}_{\bar{N}} }  \nonumber \\
\times \Bigg( [ (v_\alpha + v_\beta) \lambda_{\alpha\beta} + v_\alpha K_{\alpha\beta\alpha\alpha}
+ v_\beta K_{\alpha\beta\beta\beta} ] \bar{\chi}_{\bar{N}-1}^{[\alpha\beta]}  \nonumber \\
+ \sum_{\gamma \in \Theta, \gamma \in V}^{P_\gamma \ne P_\alpha, P_\beta} v_\gamma K_{\alpha\beta\gamma\gamma} [ \bar{\chi}_{\bar{N}-1}^{[\alpha\beta\gamma]} - (\bar{N}-1)^2 v_\alpha v_\beta \bar{\chi}_{\bar{N}-2}^{[\alpha\beta\gamma]}]  \nonumber \\
+ (\bar{N}-1)^2 (v_\alpha + v_\beta) \sum_{\gamma \in \Theta, \gamma \in V}^{P_\gamma \ne P_\alpha, P_\beta} (v_\gamma)^2 W_{\alpha\beta\gamma\gamma} \bar{\chi}_{\bar{N}-2}^{[\alpha\beta\gamma]} \Bigg) .  \label{f_VV}
\end{eqnarray}
$f_{\alpha \in V, \beta \in V}$ means $f_{\alpha\beta}$ with $\alpha \in V$, $\beta \in V$. We see that Eq. (\ref{f_VV}) is just Eq. (\ref{f_12}) applied to the $V$-subspace, while the empty $H$-subspace contributes nothing, and the full $L$-subspace changes $\epsilon_{\alpha\beta}$ to $\lambda_{\alpha\beta} = \epsilon_{\alpha\beta} + \sum_{\gamma \in L} V_{\alpha\gamma\gamma\beta}$ (monopole correction). For mixing within the $L$-subspace,
\begin{eqnarray}
f_{\alpha \in L, \beta \in L} = 0 .  \label{f_LL}
\end{eqnarray}
This is correct because the $L$-subspace is full (a determinant), so the trial wavefunction (\ref{gs_new}) and energy are invariant under mixing. For mixing within the $H$-subspace,
\begin{eqnarray}
f_{\alpha \in H, \beta \in H} = 0 .  \label{f_HH}
\end{eqnarray}
This is correct because the $H$-subspace is empty, so the trial wavefunction (\ref{gs_new}) and energy are invariant under mixing. For mixing between the $L$ and $H$ subspaces,
\begin{eqnarray}
f_{\alpha \in L,\beta \in H} = \lambda_{\alpha\beta} + \sum_{\gamma \in \Theta, \gamma \in V} W_{\alpha\beta\gamma\gamma} \langle \bar{\phi}_{\bar{N}} | \hat{n}_\gamma | \bar{\phi}_{\bar{N}} \rangle \nonumber \\
= \lambda_{\alpha\beta} + \sum_{\gamma \in \Theta, \gamma \in V} W_{\alpha\beta\gamma\gamma} (1 - \frac{ \bar{\chi}_{\bar{N}}^{[\gamma]} }{ \bar{\chi}_{\bar{N}} }) .   \label{f_LH}
\end{eqnarray}
The two forms are equivalent because of Eq. (\ref{n_ave}). We see that Eq. (\ref{f_LH}) is just the HF mean field (\ref{HF_Lcore}) corrected by the occupation numbers of the $V$-subspace $\langle \bar{\phi}_{\bar{N}} | \hat{n}_\gamma | \bar{\phi}_{\bar{N}} \rangle$ (monopole correction). For mixing between the $L$ and $V$ subspaces,
\begin{eqnarray}
f_{\alpha \in L,\beta \in V} = \frac{1}{\bar{\chi}_{\bar{N}} }   \Bigg( \lambda_{\alpha\beta} \bar{\chi}_{\bar{N}}^{[\beta]}
\nonumber \\
+ \bar{N}^2 \sum_{\gamma \in \Theta, \gamma \in V}^{P_\gamma \ne P_\beta} v_\gamma ( v_\gamma W_{\alpha\beta\gamma\gamma} - v_\beta K_{\alpha\beta\gamma\gamma} ) \bar{\chi}_{\bar{N}-1}^{[\beta\gamma]}
 \Bigg) .    \label{f_LV}
\end{eqnarray}
For mixing between the $V$ and $H$ subspaces,
\begin{eqnarray}
f_{\alpha \in V,\beta \in H} = \frac{\bar{N}^2 v_\alpha}{\bar{\chi}_{\bar{N}} }
\Bigg( v_\alpha ( \lambda_{\alpha\beta} + K_{\alpha\beta\alpha\alpha} ) \bar{\chi}_{\bar{N}-1}^{[\alpha]}  \nonumber \\
+ \sum_{\gamma \in \Theta, \gamma \in V}^{P_\gamma \ne P_\alpha} v_\gamma K_{\alpha\beta\gamma\gamma} \bar{\chi}_{\bar{N}-1}^{[\alpha\gamma]}
\nonumber \\
+ (\bar{N}-1)^2 v_\alpha \sum_{\gamma \in \Theta, \gamma \in V}^{P_\gamma \ne P_\alpha} (v_\gamma)^2 W_{\alpha\beta\gamma\gamma} \bar{\chi}_{\bar{N}-2}^{[\alpha\gamma]} \Bigg) .    \label{f_VH}
\end{eqnarray}
Because $f_{\alpha\beta} = - f_{\beta\alpha}$ is skew-symmetric, $f_{\beta \in H,\alpha \in L} = - f_{\alpha \in L,\beta \in H}$, $f_{\beta \in V,\alpha \in L} = - f_{\alpha \in L,\beta \in V}$, and $f_{\beta \in H,\alpha \in V} = - f_{\alpha \in V,\beta \in H}$ are known by Eqs. (\ref{f_LH}), (\ref{f_LV}), and (\ref{f_VH}). So we have exhausted all possible cases of $f_{\alpha\beta}$.

In this section, we show that the gradient (\ref{f_12}) becomes the expressions (\ref{f_VV})-(\ref{f_VH}) when reducing the valence space in the pairing channel. In the next section, these gradient expressions [Eqs. (\ref{f_12}) and (\ref{f_VV})-(\ref{f_VH})] are used in the iterative gradient minimizers to solve VDPC+CB.


\section{Iterative Gradient Minimizers for VDPC+CB  \label{Sec_VDPC_CB}}

The family of iterative gradient minimizers \cite{Reinhard_1982, Robledo_2011, Ring_book} is frequently used to solve the variational principle (minimize average energy). It can easily deal with one or several constraints on the solution (subsidiary conditions) \cite{Ring_book}. This is useful in many cases; for example, when we plot the potential energy surface as a function of the constrained multipole (quadrupole, octupole, $\ldots$) deformation \cite{Warda_2002}, which is needed by the generator coordinator method to go beyond the mean field \cite{Delaroche_2010}. In this section, we briefly review three gradient minimizers from the family --- the steepest descent \cite{Ring_book}, the preconditioned gradient \cite{Robledo_2011}, and ADAM \cite{Adam_2014} --- in the content of VDPC+CB. Applying them to realistic examples will be in Sec. \ref{Sec_example}.

\subsection{Steepest Descent}

VDPC+CB varies the canonical basis at fixed $v_\alpha$. We rename $\theta$ in Eq. (\ref{tran_ori}) to be $\theta_{\alpha\beta}$ (the angle mixes $|\alpha\rangle$ and $|\beta\rangle$), and introduce the vector $\vec{\theta}$ that collects all the independent $\theta_{\alpha\beta}$. So Eq. (\ref{partialE}) is rewritten as
\begin{eqnarray}
\frac{\partial \bar{E}}{\partial \theta_{\alpha\beta}} |_{\vec{\theta} = 0} = \frac{\partial (\langle \phi_N | H | \phi_N \rangle)}{\partial \theta_{\alpha\beta}} |_{\vec{\theta} = 0} = 4 f_{\alpha\beta} ,  \label{partialE_ab}
\end{eqnarray}
and its vector form is
\begin{eqnarray}
\vec{\nabla} \bar{E} = 4 \vec{f} .  \label{partialE_vec}
\end{eqnarray}
In one iteration, how to go from the current canonical basis $\{|\alpha\rangle\}$ to the new canonical basis $\{|\alpha'\rangle\}$? The steepest descent minimizer \cite{Ring_book} goes in the direction opposite to the gradient $\vec{\nabla} \bar{E}$. (The direction of the steepest descent for infinitesimal step size.) That is, we use the mixing angle
\begin{eqnarray}
\vec{\theta} = - \eta \vec{\nabla} \bar{E} = - 4 \eta \vec{f} ,  \label{theta_vec}
\end{eqnarray}
or for each component
\begin{eqnarray}
\theta_{\alpha\beta} = - 4 \eta f_{\alpha\beta} .  \label{theta_ab}
\end{eqnarray}
The parameter $\eta$ sets the step size; $\eta$ may vary from iteration to iteration \cite{Reinhard_1982, Robledo_2011}. Known $\vec{\theta}$, how to find the unitary transformation $U_{\alpha\alpha'} = \langle \alpha | \alpha' \rangle$ from $\{|\alpha\rangle\}$ to $\{|\alpha'\rangle\}$? In Eq. (\ref{theta_ab}), because $f_{\alpha\beta} = - f_{\beta\alpha}$ is a skew-symmetric matrix, we can also treat $\theta_{\alpha\beta} = - \theta_{\beta\alpha}$ as a skew-symmetric matrix. Then the matrix exponential of $\theta$
\begin{eqnarray}
U = \exp(\theta)  \label{U_exp}
\end{eqnarray}
is the desired unitary transformation $U_{\alpha\alpha'}$. (The matrix exponential of a skew-symmetric matrix is a unitary matrix.) If all the mixing angles are small, $\theta \approx 0$, the transformation (\ref{U_exp}) becomes
\begin{eqnarray}
U \approx I + \theta ,   \label{U_small}
\end{eqnarray}
which is consistent with Eq. (\ref{tran_ori}). This happens when the step size $\eta \approx 0$, or when near the energy minimum so the gradient $\vec{\nabla} \bar{E} = 4 \vec{f} \approx 0$; then $\theta \approx 0$ because of Eq. (\ref{theta_vec}) or Eq. (\ref{theta_ab}).

\subsection{Preconditioned Gradient  \label{Subsec_preGrad}}

The steepest decent minimizer performs poorly \cite{Goodfellow_book} when the Hessian matrix has a poor condition number: the partial derivative of $\bar{E}$ changes slowly in one direction (one $\theta_{\alpha\beta}$) but rapidly in another. In this case it is hard to select the step size $\eta$: $\eta$ should be small enough to avoid overshooting the energy minimum and going uphill in the direction with rapid derivative change (large positive curvature), but then the progress is too small in directions with slow derivative change (small curvature). So many iterations are needed to converge to the energy minimum.

To solve this problem, the preconditioned gradient minimizer \cite{Robledo_2011} uses different step sizes in different directions. Specifically, we introduce a precondition factor $p_{\alpha\beta}$ for each direction $\theta_{\alpha\beta}$; the vector $\vec{p}$ collects them. Then we replace Eq. (\ref{theta_vec}) by
\begin{eqnarray}
\vec{\theta} = - \eta \frac{\vec{\nabla} \bar{E}}{\vec{p}} = - 4 \eta \frac{\vec{f}}{\vec{p}} ,  \label{ptheta_vec}
\end{eqnarray}
or for each component [replacing Eq. (\ref{theta_ab}) by]
\begin{eqnarray}
\theta_{\alpha\beta} = - 4 \eta \frac{f_{\alpha\beta}}{p_{\alpha\beta}} .  \label{ptheta_ab}
\end{eqnarray}
If all $p_{\alpha\beta} = 1$, we go back to the steepest decent minimizer. Section \ref{Sec_example} will use $p_{\alpha\beta} = |e_\alpha - e_\beta| + 1 {\rm{MeV}}$, where $e_\alpha = d_\alpha / 2$ is the single-particle energy defined in Eq. (34) of Ref. \cite{Jia_2019}.


\subsection{ADAM}

ADAM means adaptive moments of the gradient \cite{Adam_2014}. It uses the decaying sum of the historical gradient and the historical (element-wise) squared gradient to compute the mixing angle $\vec{\theta}$. Specifically, ADAM keeps two vector variables $\vec{m}$ and $\vec{s}$, initialized to zero. In each iteration, we accumulate the gradient $\vec{\nabla} \bar{E} = 4 \vec{f}$ into $\vec{m}$ and $\vec{s}$ according to
\begin{eqnarray}
\vec{m} &\leftarrow& \gamma_1 \vec{m} + (1-\gamma_1) \vec{\nabla} \bar{E} ,  \label{m_moment} \\
\vec{s} &\leftarrow& \gamma_2 \vec{s} + (1-\gamma_2) (\vec{\nabla} \bar{E})^2 .  \label{s_moment}
\end{eqnarray}
In Eq. (\ref{s_moment}), $(\vec{\nabla} \bar{E})^2$ means element-wise square. The two parameters $\gamma_1$ and $\gamma_2$ control the decaying rate for weight of the historical gradient and squared gradient. In this work we use $\gamma_1 = 0.8$ and $\gamma_2 = 0.995$ (slightly smaller than their default values, $0.9$ and $0.999$, suggested in the original ADAM paper \cite{Adam_2014}). For each component, Eqs. (\ref{m_moment}) and (\ref{s_moment}) read
\begin{eqnarray}
m_{\alpha\beta} &\leftarrow& \gamma_1 m_{\alpha\beta} + 4 (1-\gamma_1) f_{\alpha\beta} ,  \label{m_moment_ab} \\
s_{\alpha\beta} &\leftarrow& \gamma_2 s_{\alpha\beta} + 16 (1-\gamma_2) (f_{\alpha\beta})^2 .  \label{s_moment_ab}
\end{eqnarray}

There is one problem about normalization \cite{Adam_2014}. At the $n$-th iteration, using Eq. (\ref{m_moment}) recursively gives
\begin{eqnarray}
\vec{m}^{(n)} &=& \gamma_1 \vec{m}^{(n-1)} + (1-\gamma_1) (\vec{\nabla} \bar{E})^{(n)}  \nonumber \\
&=& (\gamma_1)^2 \vec{m}^{(n-2)} + \gamma_1 (1-\gamma_1) (\vec{\nabla} \bar{E})^{(n-1)}  \nonumber \\
&& + (1-\gamma_1) (\vec{\nabla} \bar{E})^{(n)}  \nonumber \\
&=& \ldots  \nonumber \\
&=& (\gamma_1)^n \vec{m}^{(0)} + (1-\gamma_1) [ (\gamma_1)^{n-1} (\vec{\nabla} \bar{E})^{(1)}
\nonumber \\
&& + \ldots + \gamma_1 (\vec{\nabla} \bar{E})^{(n-1)} + (\vec{\nabla} \bar{E})^{(n)} ]  \nonumber \\
&=& (1-\gamma_1) \sum_{k = 0}^{n-1} (\gamma_1)^k (\vec{\nabla} \bar{E})^{(n-k)} ,  \nonumber
\end{eqnarray}
where $\vec{m}^{(k)}$ and $(\vec{\nabla} \bar{E})^{(k)}$ mean $\vec{m}$ and $\vec{\nabla} \bar{E}$ at the $k$-th iteration, and we have used the initial value $\vec{m}^{(0)} = 0$. So $\vec{m}^{(n)}$ is a weighted sum of the historical $(\vec{\nabla} \bar{E})^{(k)}$, $k = 1, 2, \ldots, n$, and the weights sum to
\begin{eqnarray}
(1-\gamma_1) \sum_{k = 0}^{n-1} (\gamma_1)^k = 1 - (\gamma_1)^n .  \nonumber
\end{eqnarray}
The normalized version is ($n = 1, 2, 3, \ldots$)
\begin{eqnarray}
\hat{\vec{m}}^{(n)} = \frac{\vec{m}^{(n)}}{1 - (\gamma_1)^n} .  \nonumber
\end{eqnarray}
Thus $\hat{\vec{m}}$ is the weighted {\emph{average}} of the historical gradients, because the weights sum to $1$. Similarly, the normalized version of $\vec{s}^{(n)}$ is
\begin{eqnarray}
\hat{\vec{s}}^{(n)} = \frac{\vec{s}^{(n)}}{1 - (\gamma_2)^n} .  \nonumber
\end{eqnarray}
$\hat{\vec{s}}$ is the weighted {\emph{average}} of the historical squared gradients.

In each step, ADAM replaces Eq. (\ref{theta_vec}) by
\begin{eqnarray}
\vec{\theta} = - \eta ~ \frac{\hat{\vec{m}}}{\sqrt{\hat{\vec{s}}+\delta}} ,  \label{ADAMtheta_vec}
\end{eqnarray}
or for each component [replacing Eq. (\ref{theta_ab}) by]
\begin{eqnarray}
\theta_{\alpha\beta} = - \eta ~ \frac{\hat{m}_{\alpha\beta}}{\sqrt{\hat{s}_{\alpha\beta} + \delta}} ,  \label{ADAMtheta_ab}
\end{eqnarray}
where $\delta$ is a small positive constant for numerical stabilization (to avoid dividing by zero when $\hat{s}_{\alpha\beta} \approx 0$). In this work we use $\delta = 10^{-16}$ (MeV)$^2$.

Equation (\ref{ADAMtheta_ab}) shows that ADAM uses different step sizes in different directions. If in some direction $\theta_{\alpha\beta}$, one frequently overshoots the energy minimum because of a large positive curvature, the derivatives in successive iterations will frequently change sign so cancel each other, thus $\hat{m}_{\alpha\beta}$ will be small, which will damp the step size in this direction $\theta_{\alpha\beta}$. The denominator $\sqrt{\hat{s}_{\alpha\beta}}$ in Eq. (\ref{ADAMtheta_ab}) also damps the step size in directions with large magnitude of gradient.

This section explains how to solve VDPC+CB by the family of iterative gradient minimizers. The next section explains how to solve VDPC+HFB.


\section{VDPC+HFB  \label{Sec_VDPC_HFB}}

VDPC+HFB varies $v_\alpha$ and the canonical basis together to minimize $\bar{E}$. The new algorithm accomplishes this by combining VDPC+BCS (of Ref. \cite{Jia_2019}, vary $v_\alpha$) and VDPC+CB (of Sec. \ref{Sec_VDPC_CB}, vary canonical basis). Specifically, we insert varying $v_\alpha$ (VDPC+BCS) into several places in the process of varying the canonical basis (VDPC+CB). It is unnecessary to vary $v_\alpha$ after every iteration of VDPC+CB; after every $20$ iterations, for example, is enough. In the next section, we will apply the new VDPC+HFB algorithm to a semi-realistic example.


\section{Realistic Example  \label{Sec_example}}

The semi-realistic example of the rare-earth nucleus $^{158}_{~64}$Gd$_{94}$ has been used in our recent paper \cite{Jia_2019} for VDPC+BCS. In this section, we use the same example to demonstrate the VDPC+CB and VDPC+HFB algorithm. The purpose is to show the effectiveness of the algorithms under realistic interactions, not to accurately reproduce the experimental data. For simplicity, I consider only the neutron degree of freedom, governed by the anti-symmetrized two-body Hamiltonian
\begin{eqnarray}
H = \sum_{\alpha} \epsilon_\alpha a_\alpha^\dagger a_\alpha + \frac{1}{4} \sum_{\alpha \beta \gamma \delta} V_{\alpha \beta \gamma \delta} a_\alpha^\dagger a_\beta^\dagger a_\gamma a_\delta . \label{H_example}
\end{eqnarray}
The single-particle levels $\epsilon_\alpha$ are eigenstates of the Nilsson model \cite{Nilsson_1955}. The Nilsson parameters are the same as in Ref. \cite{Jia_2017}; here I only repeat $\beta = 0.349$ (the experimental quadrupole deformation \cite{NNDC}). The neutron residual interaction $V_{\alpha \beta \gamma \delta}$ is the low-momentum {\emph{NN}} interaction $V_{{\rm{low}}{\textrm{-}}k}$ \cite{Bogner_2003} derived from the free-space N$^3$LO potential \cite{Entem_2003}.

Specifically, I use the code distributed by Hjorth-Jensen \cite{Morten_code} to compute the two-body matrix elements of $V_{{\rm{low}}{\textrm{-}}k}$ in the spherical harmonic oscillator basis up to (including) the ${\mathcal{N}} = 14$ major shell, with the standard momentum cutoff $2.1$ fm$^{-1}$. (${\mathcal{N}} = 2n_r+l$ is the major-shell quantum number.) Then the Nilsson model is diagonalized in this spherical ${\mathcal{N}} \le 14$ basis (its dimension is $D = 2 \Omega = 1360$, and $\Omega = 680$ is the number of vacancies for Kramers pairs), the eigen energies are $\epsilon_\alpha$ and the eigen wavefunctions transform the spherical two-body matrix elements into those on the Nilsson basis as used in the Hamiltonian (\ref{H_example}). The Hamiltonian (\ref{H_example}) of this work is exactly the same as that of Ref. \cite{Jia_2019} discussing VDPC+BCS.

This Hamiltonian (\ref{H_example}) has axial symmetry, so parity and angular-momentum projection onto the symmetry axis are good quantum numbers. We assume they are self-consistent symmetry; so when varying the canonical basis, the two single-particle basis states being mixed [see Eq. (\ref{tran_ori})] must have the same parity and angular-momentum projection. [The trial wavefunction $|\phi_N\rangle$ (\ref{gs}) has positive parity and zero angular-momentum projection.]

The VDPC+BCS code is written in Mathematica and runs in serial (no parallel computing, this code is from Ref. \cite{Jia_2019}). The VDPC+CB code is written in Matlab and runs in parallel. The VDPC+HFB code combines the two by calling Mathematica from Matlab. All the numerical calculations of this work were done on a laptop that has one quad-core CPU (Intel Core i7-4710MQ @ 2.5 GHz). All time costs plotted in the figures or given in the text are the actual time costs spent on this laptop. This work uses Matlab R2015a and Mathematica 10.2, to give the actual software version.


Now we discuss VDPC+CB, which varies the canonical basis at fixed $v_\alpha$. How to fix $v_\alpha$? We perform VDPC+BCS as in Ref. \cite{Jia_2019}. Specifically, we divide the full single-particle space (${\mathcal{N}} \le 14$) of dimension $1360$ ($\Omega = 680$) into $L$, $V$, $H$ subspaces. This work uses an empty $L$ subspace, a $V$ subspace of dimension $700$ that consists of all the levels having occupation number $n_\alpha > 1.63 \times 10^{-7}$, and an $H$ subspace of dimension $660$. Then we fix the $350$ variational parameters $v_\alpha$ ($350 = 700/2$) by VDPC+BCS as in Ref. \cite{Jia_2019}; the computer time cost is about $43$ seconds. Because the $V$ subspace is smaller than the full space, there is a cutoff error, which is approximately $2$ keV as can be read from Fig. 5 of Ref. \cite{Jia_2019}.

We distinguish between the energy cutoff error and the energy convergence error. It is a cutoff to divide the full single-particle space into $L$, $V$, $H$ subspaces. (Because only the $V$ subspace is active.) We define $E({\rm{final}})$ to be the converged energy (variational minimum) using a given cutoff (given $L$, $V$, $H$ subspaces). $E({\rm{exact}})$ is the converged energy in the full space (no cutoff, the $V$ subspace is the full space). The cutoff error is defined to be $E({\rm{final}}) - E({\rm{exact}})$. Using a given cutoff (given $L$, $V$, $H$ subspaces), the energy iteratively converges to $E({\rm{final}})$. The convergence error is defined to be $E({\rm{iter}}) - E({\rm{final}})$, where $E({\rm{iter}})$ is the energy after $iter$ of iterations. In the following, ``error'' means ``energy convergence error'', unless we explicitly write ``energy cutoff error''.

In passing, we list the parameters of VDPC+BCS used in this work. Skipping this paragraph does not affect reading this work; and understanding this paragraph needs Ref. \cite{Jia_2019}. This work performs VDPC+BCS similarly to Fig. 1 of Ref. \cite{Jia_2019}, but slightly changes the parameters of valence-space dimension. In step (iii), the dimension is increased to $240$ (still do $10$ iterations, more than enough for convergence). In step (v), the dimension is increased to $700$ (still do $5$ iterations, more than enough for convergence). Specifically, we sort the canonical single-particle basis states by their occupation numbers $n_\alpha$ estimated in step (iv), then from large $n_\alpha$ to small $n_\alpha$ we select $700$ basis states, which form the $V$ subspace. Because the $V$ subspace is larger, the energy cutoff error (about $2$ keV) is smaller than that in Fig. 1 of Ref. \cite{Jia_2019} (about $2.5$ keV). The same VDPC+BCS parameters ($240$, $10$ iterations; $700$, $5$ iterations) are used later in the VDPC+HFB calculations of this work. Note ``iteration'' in this paragraph means iterations in VDPC+BCS; everywhere else (outside this paragraph) ``iteration'' means iterations in VDPC+CB.

\begin{figure}
\includegraphics[width = 0.5\textwidth]{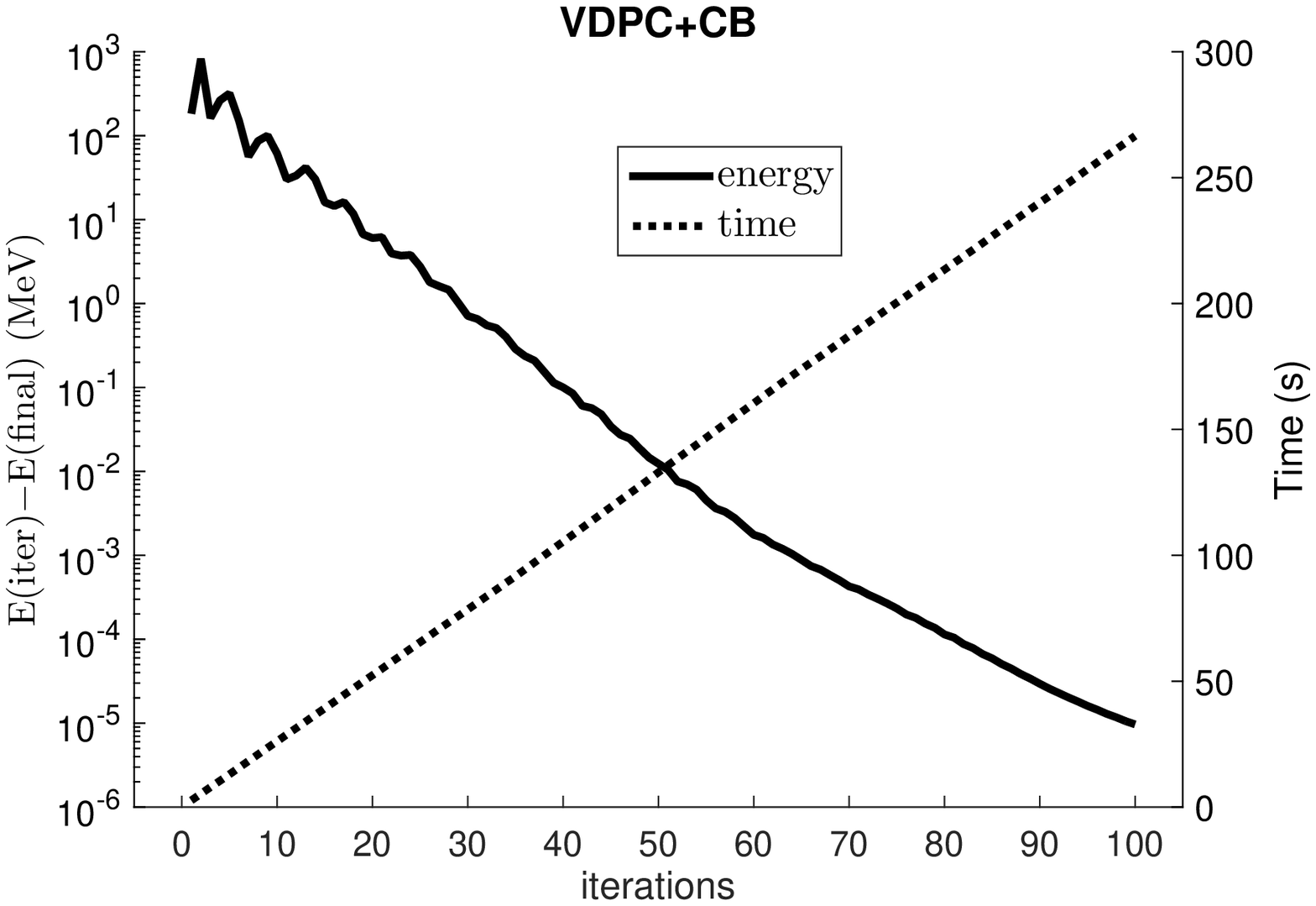}
\caption{\label{Figure_VDPC_CB} Energy and time in VDPC+CB. The horizontal axis shows the number of iterations. The solid line corresponds to the left vertical axis, and shows the energy at each iteration E(iter), relative to the final converged energy E(final). The dotted line corresponds to the right vertical axis, and shows the accumulated computer time cost after each iteration. All time costs in this work refer to that by a laptop having one quad-core CPU (Intel Core i7-4710MQ @ 2.5 GHz). }
\end{figure}

We tried to solve VDPC+CB by the three minimizers in Sec. \ref{Sec_VDPC_CB} --- the steepest descent, the preconditioned gradient, and ADAM --- and found ADAM is the most effective. (The preconditioned gradient minimizer (\ref{ptheta_ab}) uses $p_{\alpha\beta} = |e_\alpha - e_\beta| + 1 {\rm{MeV}}$. For $\alpha \in V$, $e_\alpha = d_\alpha / 2$ is the single-particle energy defined in Eq. (34) of Ref. \cite{Jia_2019}. For $\alpha \in L$ and $\alpha \in H$, $e_\alpha$ is the HF single-particle energy [including the correction by the occupation numbers of the $V$-subspace, see Eq. (\ref{f_LH})].) The steepest descent and the preconditioned gradient are conventionally used to solve HF \cite{Reinhard_1982, Robledo_2011}; for VDPC+CB they are less effective, because there could be tiny partial derivative $f_{\alpha\beta} \approx 0$ if $v_\alpha \approx v_\beta$, as shown by Eq. (\ref{f_12}). (Thus it is harder to select the step size as explained in the beginning of Sec. \ref{Subsec_preGrad}.) Figure \ref{Figure_VDPC_CB} shows the results by ADAM. Overall, the energy-error curve is linear on the log-scale plot, so energy converges exponentially with the number of iterations. The energy error drops to less than $1$ keV at the $65$th iteration, and less than $0.01$ keV at the $100$th iteration. The accumulated computer time cost increases linearly with the number of iterations, so each iteration costs the same time approximately ($2.67$ seconds in average). In Fig. \ref{Figure_VDPC_CB}, the ADAM parameters are $\gamma_1 = 0.8$, $\gamma_2 = 0.995$, $\delta = 10^{-16}$ (MeV)$^2$, and a decaying step size --- $\eta = 0.05 / 2.5^{k/100}$ at the $k$-th iteration. [See Eqs. (\ref{m_moment}), (\ref{s_moment}), and (\ref{ADAMtheta_vec}) for definitions of these ADAM parameters.] The performance (speed of convergence) of the ADAM minimizer is not very sensitive to values of these ADAM parameters.



\begin{figure}
\includegraphics[width = 0.5\textwidth]{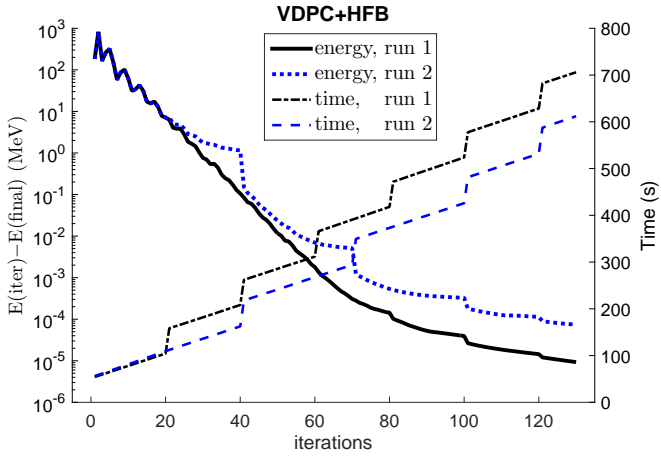}
\caption{\label{Figure_VDPC_HFB} (Color online) Energy and time in two runs of VDPC+HFB. The solid line and the dotted line correspond to the left vertical axis, and show the energy errors in the two runs. The dash-dot line and the dashed line correspond to the right vertical axis, and show the accumulated computer time cost after each iteration. (The sudden jumps in the two time-cost lines are where varying $v_\alpha$ is inserted.) }
\end{figure}

Now we discuss VDPC+HFB, which varies $v_\alpha$ and the canonical basis together. Specifically, we insert varying $v_\alpha$ (VDPC+BCS) into several places in the process of varying the canonical basis (VDPC+CB of Fig. \ref{Figure_VDPC_CB}). We ran VDPC+HFB twice, and the results are shown in Fig. \ref{Figure_VDPC_HFB}. The first run inserts varying $v_\alpha$ at (immediately after) the $0$th, $20$th, $40$th, $60$th, $80$th, $100$th, $120$th, $\ldots$ iteration (every $20$ iterations); the second run inserts varying $v_\alpha$ at the $0$th, $40$th, $70$th, $100$th, $120$th, $\ldots$ iteration. For each run, the accumulated computer time cost is a linear curve superimposed with sudden jumps where varying $v_\alpha$ is inserted. The slope of the linear curve (neglecting the jumps) is almost the same as that in Fig. \ref{Figure_VDPC_CB} (also $2.67$ seconds per iteration in average). Each jump has a similar size, around $52$ seconds, which includes the time cost of VDPC+BCS (about $43$ seconds) and some overheads. For the first run, overall the energy-error curve is linear on the log-scale plot, so energy converges
exponentially with the number of iterations. The energy error drops to less than $1$ keV at the $63$th iteration, and less than $0.1$ keV at the $82$th iteration. For the second run, varying $v_\alpha$ is inserted less frequently, waiting until the energy curve flattens out (VDPC+CB converges). In summary, the first run is more efficient than the second run if look at the two energy curves of Fig. \ref{Figure_VDPC_HFB} (energy error versus iterations); but in terms of energy error versus time cost, the two runs have similar efficiency when the energy error is bigger than $1$ keV. Comparing Fig. \ref{Figure_VDPC_CB} and Fig. \ref{Figure_VDPC_HFB}, the time cost of VDPC+HFB is roughly twice that of VDPC+CB, to achieve the same accuracy (energy error). This extra time cost of VDPC+HFB is mainly spent on the several times of varying $v_\alpha$, which is not needed in VDPC+CB. Fine-tuning the VDPC+BCS parameters can decrease this extra time.


We benchmark the speed of the VDPC algorithms against that of HF by iterative gradient minimizers. We tried to solve HF by the three minimizers in Sec. \ref{Sec_VDPC_CB} --- the steepest descent, the preconditioned gradient, and ADAM --- and found the preconditioned gradient is the most efficient (energy converges the fastest). (The preconditioned gradient minimizer (\ref{ptheta_ab}) uses $p_{\alpha\beta} = |e_\alpha - e_\beta| + 1 {\rm{MeV}}$, where $e_\alpha$ is the HF single-particle energy.)  For the Hamiltonian (\ref{H_example}), HF by the preconditioned gradient typically (we tried different particle numbers) needs $20 \sim 30$ iterations to converge within $1$ keV error, and needs $30 \sim 50$ iterations to converge within $0.01$ keV error. So HF needs less number of iterations than VDPC+CB or VDPC+HFB. On the other hand, the computer time cost per iteration in HF ($2.2$ seconds in average) is similar to that in VDPC+CB or VDPC+HFB ($2.67$ seconds in average). In each iteration, most of time is spent on basis transformation by matrix multiplication (transform the needed Hamiltonian matrix elements into the new canonical basis of the next iteration), this part is the same for HF and for VDPC+CB. The gradient expression of VDPC+CB (\ref{f_12}) is more complicated than that of HF, but this part costs insignificant computer time. In summary, based on Fig. \ref{Figure_VDPC_CB} and Fig. \ref{Figure_VDPC_HFB}, the total computer time cost of VDPC+CB and of VDPC+HFB are typically $2 \sim 3$ times and $4 \sim 6$ times that of HF, respectively, to achieve the same accuracy (energy error).

Future works can try to decrease the total computer time cost of the VDPC+HFB algorithm in two ways. First, we can decrease the time cost of VDPC+CB as suggested in Ref. \cite{Jia_2019}, for example, enable parallel computing (currently the code runs in serial). Second, we can find a better minimizer for VDPC+CB --- needs less iterations to converge. For example, we can fine-tune the ADAM parameters or introduce additional ADAM parameters (third moment, fourth moment, $\ldots$). However, it is hard to decrease the computer time cost per iteration, which is mainly spent on basis transformation by matrix multiplication (transform the needed Hamiltonian matrix elements into the new canonical basis of the next iteration).

%
%

In passing, if we do HF then VDPC+BCS, how good is the energy? (Sometimes, ``HF then BCS'' is used to approximate HFB.) For the Hamiltonian (\ref{H_example}), the (converged) HF energy is $1.225$ MeV higher than the (converged) VDPC+HFB energy. VDPC+BCS (after HF) further lowers the energy by $0.326$ MeV, so the (converged) energy of ``HF then VDPC+BCS'' is still $0.899$ MeV higher than VDPC+HFB. (The pairing energy $0.326$ MeV is small, because the density of the HF single-particle levels is accidentally small near the Fermi surface.) Therefore, ``HF then VDPC+BCS'' is not good enough, and VDPC+HFB is necessary.



\section{Conclusions  \label{Sec_summary}}

Recently Ref. \cite{Jia_2019} proposed a scheme that applies the variational principle directly to the coherent pair condensate in the BCS case (VDPC+BCS). This work extends the scheme to the HFB case (VDPC+HFB). The result is equivalent to that of the so-called variation after particle-number projection in the HFB case (VAP+HFB), but now the particle number is always conserved and the time-consuming projection is avoided. The HFB theory is frequently criticized for breaking the exact particle number. Meanwhile, VAP+HFB by the numerical gauge-angle integration may not be very easy, and in the literature there are far fewer realistic applications of VAP+HFB than those of HFB without projection. We hope the new VDPC+HFB algorithm could become a common practice because of its simplicity.

Specifically, this work derives the analytical expression for the gradient of the average energy with respect to changes of the canonical basis. The VDPC+CB algorithm supplies this gradient expression to the family of gradient minimizers to iteratively minimize energy. In practice, we find the so-called ADAM minimizer, borrowed from the machine-learning field, is very effective. The VDPC+HFB algorithm combines VDPC+BCS and VDPC+CB, by inserting VDPC+BCS (vary $v_\alpha$) into several places in the process of VDPC+CB (vary the canonical basis). The new algorithms are demonstrated in a semi-realistic example using the realistic $V_{{\rm{low}}{\textrm{-}}k}$ interaction and large model spaces (up to $15$ harmonic-oscillator major shells). They easily run on a laptop, and practically the computer time cost to solve VDPC+HFB (time to solve VDPC+CB) is about $4 \sim 6$ times ($2 \sim 3$ times) that to solve HF by the iterative gradient minimizers. Future works can further optimize the code for less time cost, as discussed in Sec. \ref{Sec_example}.

The parameters in nuclear mean-field interactions \cite{Bender_2003} are usually fitted (to experimental data) in the theory of HF, HFB, or approximate VAP+HFB (for example, the Lipkin-Nogami prescription followed by particle-number projection \cite{Samyn_2004}). These fitting parameters should be fine-tuned to generate optimum interactions for VDPC+HFB. In addition, higher-order correlations beyond VDPC can be included, for example, by the generalized-seniority truncation of the shell model \cite{Talmi_book, Allaart_1988, Zhao_2014, Jia_2015, Jia_2016_ph, Jia_2016_Sn, Qi_2016, Jia_2017}.

\section{Acknowledgements}

Support is acknowledged from the National Natural Science
Foundation of China No. 11405109.

\newpage
~
\newpage

\end{document}